\documentclass[prl,superscriptaddress,twocolumn,amsmath]{revtex4-1}

\usepackage{bm}
\usepackage{graphicx}
\usepackage[usenames,dvipsnames,svgnames,table]{xcolor}
\usepackage[colorlinks=true,linkcolor=RoyalBlue,citecolor=RoyalBlue]{hyperref}

\begin{document}

\title{Anomalous circular phonon dichroism in transition metal dichalcogenides}

\author{Wen-Yu Shan}
\affiliation{Department of Physics, School of Physics and Materials Science, Guangzhou University, Guangzhou 510006, China}

\date{\today}

\begin{abstract}
  Magnetic field can generally induce circular phonon dichroism based on the formation of Landau levels of electrons. Here we study the magnetization-induced circular phonon dichroism in transition metal dichalcogenides, without forming the Landau levels. We find that, instead of the conventional deformation potential coupling, the pseudogauge-type electron-phonon coupling plays an essential role in the emergence of the phenomenon. As a concrete example, a large dichroism signal is obtained in monolayer MoTe$_2$ on a EuO substrate, even without considering the Rashba spin-orbit coupling. Due to the two-dimensional spin-valley-coupled band structure, MoTe$_2$ shows a reciprocal and nonreciprocal absorption of circularly polarized acoustic phonons upon reversing the direction of phonon propagation and magnetization, respectively. By varying the gate voltage, a tunable circular phonon dichroism can be realized, which paves a way toward new physics and applications of two-dimensional acoustoelectronics.

\end{abstract}

\maketitle

$\textit{Introduction}$.---Recent years have seen a surge of interest in investigating topological properties in the nonelectronic systems, e.g., photonic, magnonic and phononic materials. For phonons, the concepts of band topology and geometry have brought into new ingredients: chiral phonons~\cite{zhang2015,zhu2018,chen2019,maity2021}, angular momentum~\cite{zhang2014,hamada2018,zhang2021,komiyama2021}, orbital magnetic moments of phonons~\cite{juraschek2017,juraschek2019,cheng2020,juraschek2020}, phonon angular momentum Hall effect~\cite{park2020}, phonon rotoelectric effect~\cite{hamada2020} and so on. In metals, the interplay between phonons and electrons with nontrivial band topology or geometry may further induce distinctive features, such as phonon helicity~\cite{hu2021} and phonon magnetochiral effect~\cite{nomura2019,sengupta2020}.

\begin{figure}[t]
\centering \includegraphics[width=0.39\textwidth]{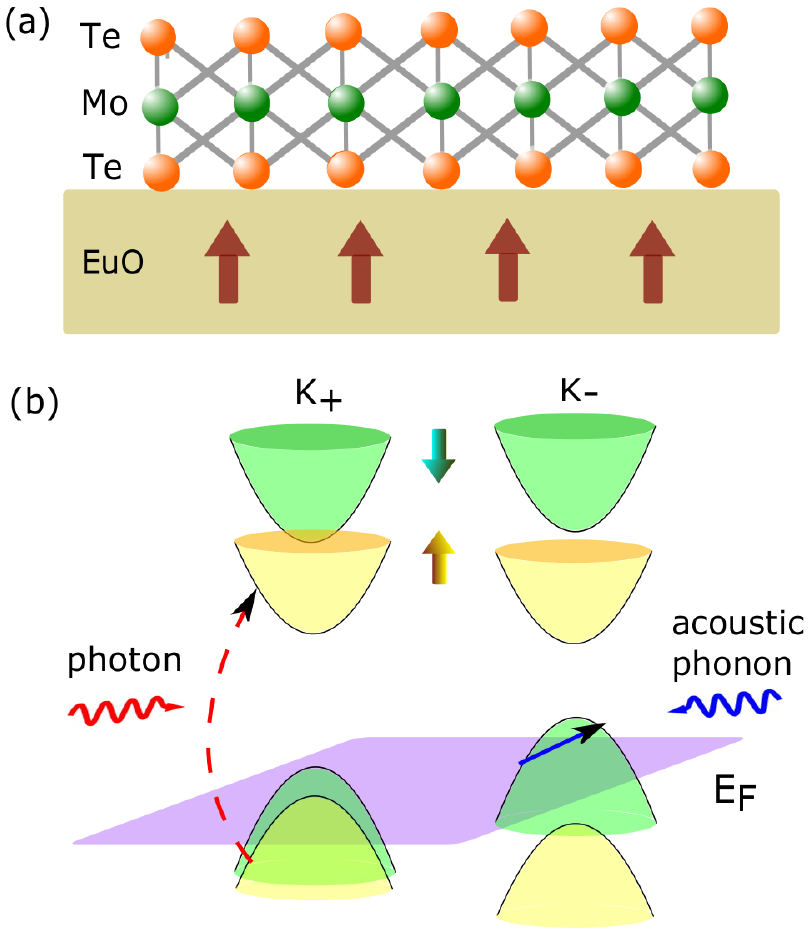}
\caption{ Schematics of (a) the setup and (b) electronic band structure of monolayer MoTe$_2$. In (a), 2H-phase monolayer MoTe$_2$ is deposited on the EuO substrate. In (b), yellow (green) region corresponds to spin-up (-down) bands. Transition process of electrons due to acoustic phonons (photons) is indicated by the blue solid (red dashed) line.}
\label{fig:cpd_setup}
\end{figure}

Circular dichroism, the differential absorption between left- and right-handed circularly polarized light, has been widely used in examining topological phases of matter~\cite{souza2008,wang2011,tran2017,liu2018,pozo2019,repellin2019,schuler2020}. A phononic analog, namely, circular phonon dichroism (CPD), is later proposed in three-dimensional Weyl semimetals~\cite{liu2017}. However, a direct analogy between phonons and photons is not that obvious. The reasons are twofold. First, the photon wave vector is usually much smaller than the Fermi wave vector of electrons, thus only inducing the interband transition of electrons; whereas the phonon wave vector may be comparable to that of electrons, giving rise to either interband or intraband transition (see Fig. \ref{fig:cpd_setup} (b)). Second, light waves consist of only transverse modes, whereas acoustic waves in solids have both longitudinal and transverse modes. Particularly, when dealing with two-dimensional (2D) materials, one has to mix longitudinal and transverse in-plane modes to create circular phonons~\cite{sonntag2021}, in marked contrast to the case of light. This indicates that 2D circular phonon dichroism is intrinsically different from the circular dichroism of light, where the former has received far less attention.

Experimentally, several works have unveiled the effect of Landau levels of electrons on the phonon dispersion or circular dichroism in graphene~\cite{kossacki2012,kumaravadivel2019,sonntag2021}, such as the magnetophonon resonance. Nevertheless, the treatment of Landau levels inevitably induces topology, even into an originally trivial system. In this sense, the CPD can not resolve the real band topology or geometry of the underlying system. Another way of breaking time-reversal symmetry is to introduce the magnetic exchange interaction, which does not require the formation of Landau levels and could retain the basic topology or geometry of the band structure. Up to now, the intrinsic magnetization-induced CPD in 2D materials like monolayer transition metal dichalcogenides, remains unknown. This generalization of magnetization bears similarities to the case of anomalous Hall effect, hence the name $anomalous$ $circular$ $phonon$ $dichroism$. The distinct spin-valley-coupled band structure of transition metal dichalcogenides may further contribute to the anomalous behaviors of CPD and their nonreciprocal relations. Therefore studying this new type of CPD would be desirable for a better understanding and manipulation of band geometry or topology in 2D materials.

In this paper, we explore the magnetization-induced CPD in monolayer transition metal dichalcogenides. To allow this effect, the pseudogauge-type electron-phonon coupling is necessary instead of the conventional deformation potential coupling. We obtain a large dichroism signal in monolayer MoTe$_2$ on a EuO substrate, even in the absence of Rashba spin-orbit coupling. Due to the unique spin-valley coupling, we find that MoTe$_2$ shows a reciprocal (nonreciprocal) absorption of circularly polarized acoustic phonons upon reversing the direction of phonon propagation (magnetization). Our study refreshes our knowledge on the effect of electron-phonon coupling on phonon dynamics, and paves the way toward acoustoelectronics for 2D materials.

$\textit{Model Hamiltonian}$.---We take the pristine 2H-phase transition metal dichalcogenides MoTe$_2$ on a EuO substrate as a prototype (see Fig. \ref{fig:cpd_setup} (a)). The effective electronic Hamiltonian is given by $\mathcal{H}_e=\sum_{\bm k}\psi^+(\bm k)[H_0+H_{soc}+H_{ex}+H_R]\psi(\bm k)$, where~\cite{qi2015,habe2017}
\begin{equation}
\begin{split}\label{Hamiltonian}
H_0 &= \hbar v(\tau\sigma_xk_x+\sigma_yk_y) + \frac{\Delta}{2}\sigma_z,\\
H_{soc} &= \tau s_z (\lambda_c\sigma_++\lambda_v\sigma_-),\\
H_{ex} &= -\bm s\cdot\bm n(B_c\sigma_++B_v\sigma_-),\\
H_R &= \lambda_R (\tau s_y\sigma_x-s_x\sigma_y).
\end{split}
\end{equation}
$\psi^+(\bm k)$ and $\psi(\bm k)$ are the creation and annihilation operator of electrons. $H_{soc}$, $H_{ex}$ and $H_R$ correspond to the Ising-type spin-orbit coupling, proximity-induced exchange and Rashba interaction, respectively. $\bm s$ and $\bm \sigma$ are Pauli matrices acting on spin $\{\uparrow,\downarrow\}$ and orbit subspace $\{|d_{z^2}\rangle,\frac{1}{\sqrt{2}}(|d_{x^2-y^2}\rangle+i\tau|d_{xy}\rangle)\}$, and $\sigma_{\pm}=\frac{1}{2}(\sigma_0\pm\sigma_z)$. $\tau=\pm1$ labels valley $K_{\pm}$. $\lambda_{c/v}$ describes the spin splitting of the conduction and valence bands, respectively. $B_{c/v}$ is the effective Zeeman field experienced by the conduction and valence bands, arising from the exchange coupling with the magnetic substrate. The out-of-plane $z$-direction magnetization $\bm n=\bm e_z$ is considered (see Fig. \ref{fig:cpd_setup} (a)). For the moment, we set $\lambda_R=0$ in order to have analytical expressions and intuitive physical picture. The role of $\lambda_R$ will be clarified later. The electronic band structure upon magnetization is schematically shown in Fig. ~\ref{fig:cpd_setup} (b), where the signature of spin-valley coupling can be seen explicitly. The Fermi level is pinned at the valence bands, where the effect of spin-valley coupling is manifest. The magnetization $B_{c/v}$ shifts the opposite-spin states from different valleys in opposite directions, and thus breaks the time-reversal symmetry.

For the phononic part, we consider two branches of in-plane acoustic phonon modes. Due to the low sound velocity $c_{l/t}$ ($l/t$ for longitudinal/transverse phonon polarization), the acoustic phonon energy $\omega_{l/t}=\hbar c_{l/t}|\bm q|$ is much smaller than the valence band splitting of electrons, i.e., $2(B_v\pm\lambda_v)$, where $\bm q$ is the phonon wave vector. As a result, only intraband transitions of electrons are triggered by acoustic phonons (see Fig. ~\ref{fig:cpd_setup} (b)). By contrast, optical phonon modes with larger energy, may enable either intraband or interband transitions. Nonetheless, the basic physical picture should be similar. For simplicity, we further study the long-wavelength limit of phonon modes, which allows us to neglect the intervalley scattering process of electrons.

Based on the theory of elasticity~\cite{landau1959,suzuura2002,cazalilla2014}, the electron-acoustic-phonon coupling in MoTe$_2$ contains two terms: $\mathcal{H}_{e-ph}=\mathcal{H}_{e-ph}^{d}+\mathcal{H}_{e-ph}^{p}$, where $\mathcal{H}_{e-ph}^{d}$ ($\mathcal{H}_{e-ph}^{p}$) refers to the deformation (pseudogauge) potential coupling. $\mathcal{H}_{e-ph}^{d/p}$ has a general form~\cite{shan2020,supple}
\begin{equation}
\begin{split}
\mathcal{H}_{e-ph}^{d/p} &= \sum_{\bm k,\bm q}\psi^+(\bm k+\bm q)[\bm u(\bm q)\cdot\hat{\bm T}_{d/p}(\bm q)]\psi(\bm k),
\end{split}
\end{equation}
where $\bm u(\bm q)$ is the Fourier transform of the in-plane collective displacement $\bm u(\bm r)$ for acoustic modes~\cite{suzuura2002} and $\hat{\bm T}(\bm q)$ is the Fourier transform of the ``effective" force operator $\hat{\bm T}(\bm r)$ acting on atoms by electrons. For the deformation potential $\mathcal{H}_{e-ph}^{d}$, $\hat{\bm T}_d(\bm q)=ig_d\bm q$, which is independent of the valley index $\tau$. For the pseudogauge potential $\mathcal{H}_{e-ph}^{p}$, the force operator becomes valley-dependent, that is, $\hat{\bm T}^{\tau=-1}_p(\bm q)=ig_p[\bm q\cdot\bm\sigma,(\bm q\times\bm\sigma)_z]$ and $\hat{\bm T}^{\tau=1}_p(\bm q)=\mathcal{K}[\hat{\bm T}^{\tau=-1}_p(-\bm q)]$. $\mathcal{K}$ is the complex conjugation operator. The relation between $\hat{\bm T}_{p}^{\tau=1}(\bm q)$ and $\hat{\bm T}_{p}^{\tau=-1}(-\bm q)$ preserves the time-reversal symmetry of electron-phonon coupling in the absence of magnetization.

$\textit{Phonon equation of motion}$.---For the phonon dynamics, we consider the phonon equation of motion in the frequency-momentum ($\omega$, $\bm q$) domain~\cite{liu2017} 
\begin{equation}
\omega^2u_{\alpha}(\bm q)=\sum_{\beta}[\Phi_{\alpha\beta}(\bm q)
+\hbar\chi_{\alpha\beta}(\bm q,\omega)]u_{\beta}(\bm q), \;
\end{equation}
where $\alpha,\beta=x,y$ and $\Phi(\bm q)$ is the dynamical matrix. $\chi_{\alpha\beta}(\bm q,\omega)$ is a retarded response function arising from the electron-phonon coupling and follows at each valley~\cite{supple}
\begin{equation}
\begin{split}
&\chi_{\alpha\beta}^{\tau}(\bm q,\omega+i\delta)=\sum_{n,m}
\int\frac{\hbar d^2\bm k}{\rho(2\pi)^2}
\frac{f_{\tau,m,\bm k}-f_{\tau,n,\bm k-\bm q}}{\omega+i\delta+E_{\tau,m,\bm k}-E_{\tau,n,\bm k-\bm q}}\\
&\times\langle \tau, m,\bm k|\hat{T}_{\alpha}^{\tau}(\bm q)|\tau, n,\bm k-\bm q\rangle
\langle \tau, n,\bm k-\bm q|\hat{T}_{\beta}^{\tau}(-\bm q)|\tau, m,\bm k\rangle.
\end{split}
\end{equation}
$E_{\tau,m,\bm k}$ and $|\tau,m,\bm k\rangle$ are the dispersion and electronic wave function of Hamiltonian (\ref{Hamiltonian}), respectively. $f_{\tau,m,\bm k}$ ($f_{\tau,n,\bm k-\bm q}$) is the Fermi distribution function, $\rho$ is the 2D mass density, $\delta$ is a positive infinitesimal. Since only intraband transitions (band indices $m=n$) of electrons are allowed by acoustic modes in the low-temperature limit, $m,n$ reduce to the ones intersected by the Fermi level, i.e., spin-split valence bands at valley $K_{\pm}$ (see Fig.~\ref{fig:cpd_setup} (b)).

$\textit{Circular phonon dichroism}$.---Our main interest lies in the anti-Hermitian part of $\chi(\bm q,\omega)$, that is, $-2i\omega\gamma(\bm q,\omega)$, where $\gamma(\bm q,\omega)$ is a Hermitian matrix satisfying $\gamma^+(\bm q,\omega)=\gamma(\bm q,\omega)$. This matrix corresponds to the non-Hermitian part of the phonon self-energy, which physically originates from the phonon absorption by electrons. In the basis of $\{\hat{x},\hat{y}\}^T$, $\gamma$ matrix has the form
\begin{equation}
\begin{split}
\gamma(\bm q,\omega) = 
\left[\begin{array}{cc}
D(\bm q,\omega)+\bar{D}(\bm q,\omega)   &   \bar{A}(\bm q,\omega)+iA(\bm q,\omega)  \\
\\
\bar{A}(\bm q,\omega)-iA(\bm q,\omega)  &   D(\bm q,\omega)-\bar{D}(\bm q,\omega)  \\
\end{array}\right].
\end{split}
\end{equation}
Different from the Weyl semimetals~\cite{liu2017}, new terms $\bar{D}(\bm q,\omega)$ and $\bar{A}(\bm q,\omega)$ occur in monolayer MoTe$_2$ as a result of $D_{3h}$ point-group symmetry. For the left- and right-handed circularly polarized phonons, $|u_{L/R}\rangle=$ $\frac{1}{\sqrt{2}}[1$ $\pm i]^T$, the damping (absorption) coefficients read $\gamma^{L/R}=$ $D(\bm q,\omega)\mp A(\bm q,\omega)$. The relative difference between $\gamma^{L}$ and $\gamma^{R}$ defines the $circular$ $phonon$ $dichroism$ (CPD). One can see that the behavior of CPD is totally determined by $A(\bm q,\omega)/D(\bm q,\omega)$.

For longitudinal or transverse phonons, the polarization is linear as $|u_l\rangle=$ $[\cos\phi_{\bm q}$ $\sin\phi_{\bm q}]^T$ and $|u_t\rangle=$ $[-\sin\phi_{\bm q}$ $\cos\phi_{\bm q}]^T$, where the angular variable $\phi_{\bm q}=\tan^{-1}(q_y/q_x)$. The damping coefficients are given by $\gamma^{l/t}=$ $D(\bm q,\omega)$ $\pm$ $\cos2\phi_{\bm q}\bar{D}(\bm q,\omega)$ $\pm$ $\sin2\phi_{\bm q}\bar{A}(\bm q,\omega)$, which explicitly depends on the phonon propagation direction $\bm q$. Here, different from the circular phonons, the damping coefficients $\gamma^{l/t}$ for the linear phonons depend on the parameters $\bar{D}(\bm q,\omega)$ and $\bar{A}(\bm q,\omega)$.

Specifically for the deformation potential $\mathcal{H}_{e-ph}^{d}$, $\gamma$ matrix is proportional to~\cite{supple}
\begin{equation}
\begin{split}
\gamma(\bm q,\omega) \propto 
\left[\begin{array}{cc}
q_x^2  &   q_xq_y  \\
\\
q_xq_y  &   q_y^2  \\
\end{array}\right].
\end{split}
\end{equation}
This immediately leads to $A(\bm q,\omega)=0$, meaning that the CPD vanishes when only the deformation potential coupling is taken in account. Meanwhile, $\gamma^t=0$, suggesting that there is no absorption for the transverse phonon modes. This agrees with the fact that the deformation potential only couples electrons to the longitudinal phonon modes~\cite{shan2020}.

\begin{figure}[t]
\centering \includegraphics[width=0.49\textwidth]{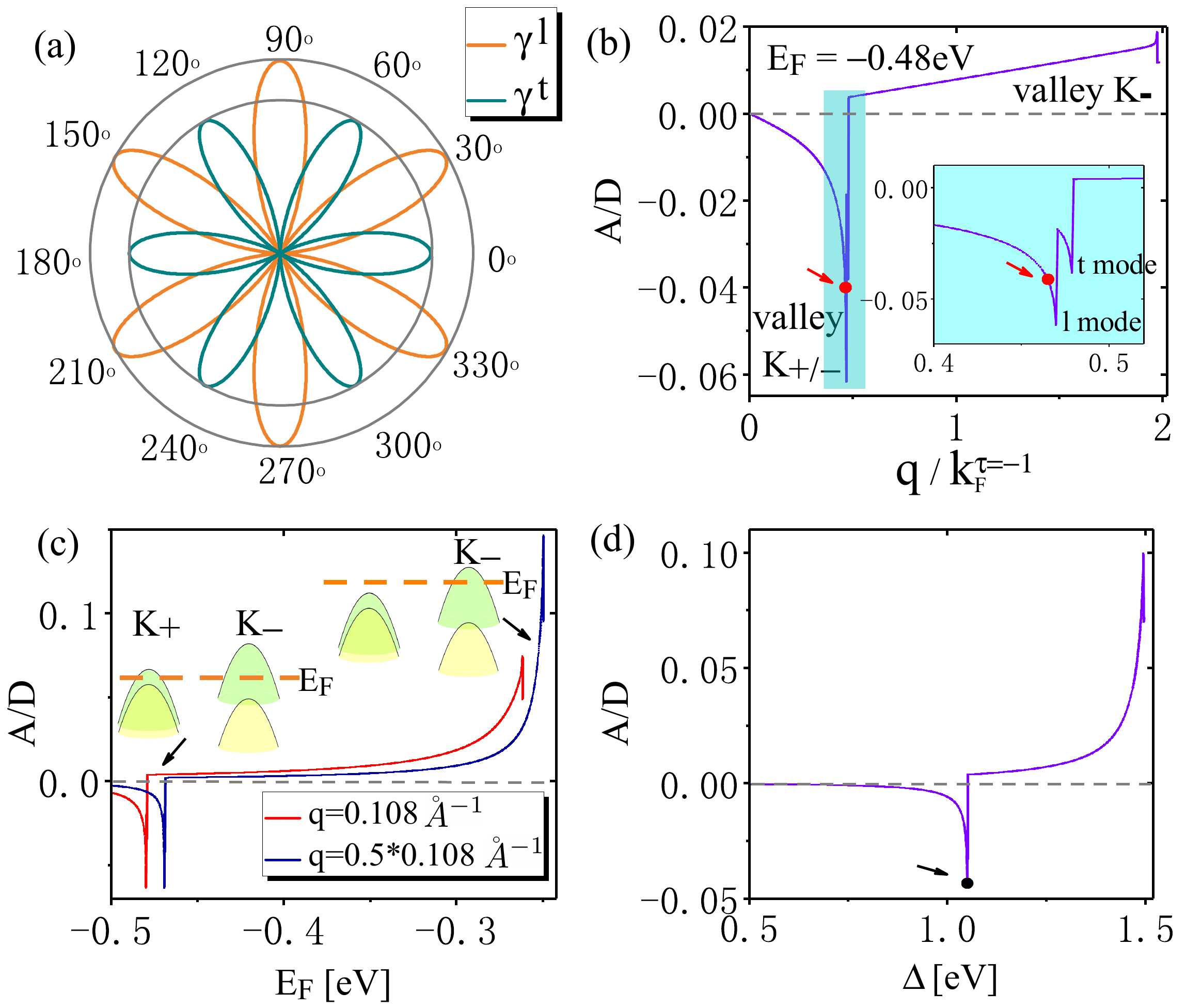}
\caption{ (a) Angular $\phi_{\bm q}$-dependence of the damping coefficients $\gamma^l$ ($\gamma^t$) for the longitudinal (transverse) acoustic phonon modes. (b)-(d) Relations of the circular phonon dichroism $A/D$ versus the phonon wave vector $q/k_F^{\tau=-1}$, Fermi energy $E_F$ and gap function $\Delta$, respectively. In (b), the Fermi energy is fixed: $E_F=-0.48$ eV.  The inset shows the details of the cyan region, and the two peaks are due to the longitudinal and transverse phonon modes, respectively. The peaks in the cyan region are given by a summation of valleys $K_{\pm}$, whereas the peaks near $q/k_F^{\tau=-1}=2$ are only determined by valley $K_-$. $k_F^{\tau=-1}$ is the Fermi wave vector at valley $K_-$. The red dot refers to the case of (a). In (c), different values of $q$ are adopted. The locations of the Fermi level at the peaks are shown in the inset: both valley $K_{\pm}$ are intersected at smaller $E_F$; only valley $K_-$ is intersected at larger $E_F$. In (d), the black dot indicates the value of $\Delta$ in (a)-(c): $\Delta=1.05$ eV. Parameters: $\lambda_v=0.11$ eV, $\lambda_c=0.029$ eV, $\hbar v=2.33$ eV$\cdot\AA$, $B_c=0.206$ eV, $B_v=0.17$ eV~\cite{qi2015}, longitudinal and transverse sound velocity $c_l=3.64\times10^3$ m/s and $c_t=2.21\times10^3$ m/s~\cite{rano2020}, mass density $\rho=9.40\times10^{-6}$ kg/m$^2$~\cite{keum2015} and the electron-phonon coupling constant $g_p=0.32$ eV~\cite{chen2016}. }
\label{fig:cpd_mote2}
\end{figure}

For the pseudogauge potential $\mathcal{H}_{e-ph}^{p}$, the situation is more complex. When only focusing on the acoustic modes, analytical expressions for all elements of $\gamma(\bm q,\omega)$ matrix can be obtained~\cite{supple}. For example, when a single valence band at valley $K_{\tau}$ is intersected by the Fermi level, $A(\bm q,\omega)/D(\bm q,\omega)$ reduces to
\begin{equation}
\begin{split}\label{analytic}
\frac{A^{\tau}(\bm q,\omega)}{D^{\tau}(\bm q,\omega)}
&=\tau\omega\frac{\Delta-\tau\lambda_c+\tau\lambda_v+B_c-B_v}{2[\omega x_F^{\tau}-(\hbar vk_F^{\tau})^2]}
\Theta(k_F^{\tau}-\frac{q}{2}-k_0^{\tau}),
\end{split}
\end{equation}
where the Heaviside step function $\Theta(\cdots)$ constrains the magnitude of phonon wave vector $q=|\bm q|$. $k_F^{\tau}$ is a valley-dependent Fermi wave vector of electrons. $k_0^{\tau}=$ $\frac{\omega}{2\hbar v}\sqrt{1+\frac{(\Delta-\tau\lambda_c+\tau\lambda_v+B_c-B_v)^2}{(\hbar vq)^2-\omega^2}}$ and $x_F^{\tau}=$ $\sqrt{(\frac{\Delta-\tau\lambda_c+\tau\lambda_v+B_c-B_v}{2})^2+(\hbar vk_F^{\tau})^2}$. However, such a simple relation fails when both valleys are intersected by the Fermi level, given that $A(\bm q,\omega)=\sum_{\tau}A^{\tau}(\bm q,\omega)$ and $D(\bm q,\omega)=\sum_{\tau}D^{\tau}(\bm q,\omega)$. On the other hand, both $\bar{A}(\bm q,\omega)$ and $\bar{D}(\bm q,\omega)$ become $\phi_{\bm q}$-dependent~\cite{supple}: $\bar{A}(\bm q,\omega)=-F(q,\omega)\sin4\phi_{\bm q}$ and $\bar{D}(\bm q,\omega)=F(q,\omega)\cos4\phi_{\bm q}$, with a $\phi_{\bm q}$-independent factor $F(q,\omega)$. By substituting these into $\gamma^{l/t}$, we find for linearly polarized phonons,
\begin{equation}
\begin{split}
\gamma^{l/t}=D(q,\omega)\pm F(q,\omega)\cos6\phi_{\bm q}.
\end{split}
\end{equation}
One can see that $\gamma^{l/t}$ has a six-fold ($C_6$) rotational symmetry on $\phi_{\bm q}$ (see Fig. \ref{fig:cpd_mote2} (a)), which is different from the three-fold ($C_3$) rotational symmetry of the underlying crystals. The reason for the symmetry mismatch is due to the reciprocal behaviors of $\gamma^{l/t}$ upon reversing the direction of phonon propagation, i.e., $\bm q\rightarrow-\bm q$, as shown in Table \ref{tab:nonreciprocity}. For phonons, $c_{l/t}\ll v$, giving rise to~\cite{supple} $D(q,\omega)\approx -F(q,\omega)$. As a result, $\gamma^{l}\approx2D(q,\omega_l)\sin^23\phi_{\bm q}$ and $\gamma^{t}\approx2D(q,\omega_t)\sin^23(\phi_{\bm q}-\frac{\pi}{6})$. This means that there is an angular shift $\frac{\pi}{6}$ in $\phi_{\bm q}$ between $\gamma^l$ and $\gamma^t$, as shown in Fig. \ref{fig:cpd_mote2} (a).

\renewcommand\arraystretch{1.5}

\begin{table}[bp]
\caption{ Transformation properties of parameters $D$, $\bar{D}$, $A$ and $\bar{A}$.}
\label{tab:nonreciprocity}%
\begin{ruledtabular}
\begin{tabular}{*5c} 
Transformation & $D(\bm q,\omega)$  & $\bar{D}(\bm q,\omega)$ & $A(\bm q,\omega)$ & $\bar{A}(\bm q,\omega)$  \\  \hline
$\bm q\rightarrow-\bm q$ & $+$ & $+$ & $+$ & $+$ \\ 
$\bm n\rightarrow-\bm n$ & $+$ & $+$ & $-$ & $+$\\
\end{tabular}
\end{ruledtabular}
\end{table}

For circularly polarized phonons, numerical results of $A(\bm q,\omega)/D(\bm q,\omega)$ as functions of the rescaled phonon wave vector $q/k_F^{\tau=-1}$, Fermi energy $E_F$ and $\Delta$ are shown in Fig. \ref{fig:cpd_mote2} (b)-(d), respectively. The Fermi wave vector $k_F^{\tau=-1}$ rather than $k_F^{\tau=1}$ is selected since the valence band edge of valley $K_-$ is higher than $K_+$, as shown in Fig. \ref{fig:cpd_setup}. In Fig. \ref{fig:cpd_mote2} (b), a non-monotonic behavior of $A/D$ as $q$ increases can be seen explicitly. The jumps at $q/k_F^{\tau=-1}\approx0.47$ and $1.98$ originate from the sudden vanishing of valley $K_+$ and $K_-$, respectively, as required by the factor $\Theta(k_F^{\tau}-\frac{q}{2}-k_0^{\tau})$ in Eq. (\ref{analytic}). Such a factor can be understood as a result of the energy and momentum conservation for the electron-phonon scattering process. For acoustic phonons, the electron scattering approximately occurs on the Fermi surface. In this sense, the phonon wave vector $q$ must be smaller than the maximum value of momentum transfer of electrons, that is, $q<2k_F^{\tau}$. $k_0^{\tau}$ is a small offset wave vector arising from the acoustic phonon dispersion $\omega$. As seen in the inset of Fig. \ref{fig:cpd_mote2} (b), there are actually two adjacent peaks (jumps) in the highlighted region corresponding to the $l$ and $t$ mode, respectively, since $k_0^{\tau}$ is different for $\omega=\omega_{l/t}$. As the sound velocity $c_l>c_t$, $k_0^{\tau}$ is larger for the longitudinal mode, leading to a smaller transition value of $q$. In Fig. \ref{fig:cpd_mote2} (c), the locations of the Fermi level for the peaks are indicated in the inset. The peaks at the lower (higher) Fermi level are dominated by valley $K_+$ ($K_-$), which exhibit opposite signs of $A/D$. For each valley, the magnitude $|A/D|$ increases when the Fermi level is tuned toward the band edge. Different values of $q$ are also compared. We find that by adopting a smaller $q$, the peaks are shifted to a higher Fermi level, as $k_F^{\tau}$ becomes smaller. The peaks also show a larger magnitude and become sharper, particularly for the second peaks. Therefore this provides a means of tuning the sign and magnitude of the CPD. In Fig. \ref{fig:cpd_mote2} (d), the value of $\Delta$ adopted in Fig. \ref{fig:cpd_mote2} (a)-(c) is indicated. We can see that the magnitude $|A/D|$ is basically enhanced when $\Delta$ increases, expect for the discontinuous points. That is the reason why we propose monolayer transition metal dichalcogenides as candidate materials, which have large band gap and thus large CPD signals. For an order-of-magnitude estimate, we consider the parameters corresponding to the red dot in Fig. \ref{fig:cpd_mote2} (b), which also refer to the case of Fig. \ref{fig:cpd_mote2} (a). We find $D=1.90\times10^7$/s and $A=-7.81\times10^5$/s. This yields a difference of the attenuation between the left- and right-handed circularly polarized waves, that is, $(\gamma^L-\gamma^R)/\bar{c}\sim534$/m, where $\bar{c}=(c_l+c_t)/2$ is the average sound velocity. Such difference is much larger than that of the Weyl semimetals~\cite{liu2017}, and should be observable in ultrasonic experiments.

$\textit{Nonreciprocal absorption}$.---Given that both the space-inversion and time-reversal symmetry are broken in our system, the absorption of circularly polarized phonons is expected to be nonreciprocal. To see this, we consider in Table \ref{tab:nonreciprocity} the transformation properties of parameters $D$, $\bar{D}$, $A$ and $\bar{A}$ upon reversing the direction of phonon propagation $\bm q$ or magnetization $\bm n$. We find that $D$, $\bar{D}$ and $\bar{A}$ are even functions of $\bm q$ and $\bm n$, whereas $A$ is an even (odd) function of $\bm q$ ($\bm n$). Accordingly, the absorption coefficients of circular phonons $\gamma^{L/R}$ remain unchanged under the transformation $\bm q\rightarrow-\bm q$, whereas $\gamma^{L/R}$ interchange with each other under the transformation $\bm n\rightarrow-\bm n$. This represents a reciprocal and nonreciprocal CPD upon reversing the direction of phonon propagation and magnetization, respectively. Such result is similar to that of the Faraday rotation of light polarization~\cite{freiser19682002}, where the rotation angle only depends on the magnetic field direction. However, the origin is different. The absorption of circularly polarized photons is actually nonreciprocal when $\bm q\rightarrow-\bm q$, but the chirality of circular photons also depends on the light propagation direction $\bm q$. As a result, the reciprocal $\bm q$-dependence of the rotation angle is recovered. On the other hand, due to the 2D nature, the chirality of in-plane circular phonons is independent of the phonon propagation direction $\bm q$, giving rise to the reciprocal absorption. This also indicates that there is no directional dichroism~\cite{fuchs1965} or phonon magnetochiral effect~\cite{nomura2019,sengupta2020} in our system.

\begin{figure}[t]
\centering \includegraphics[width=0.47\textwidth]{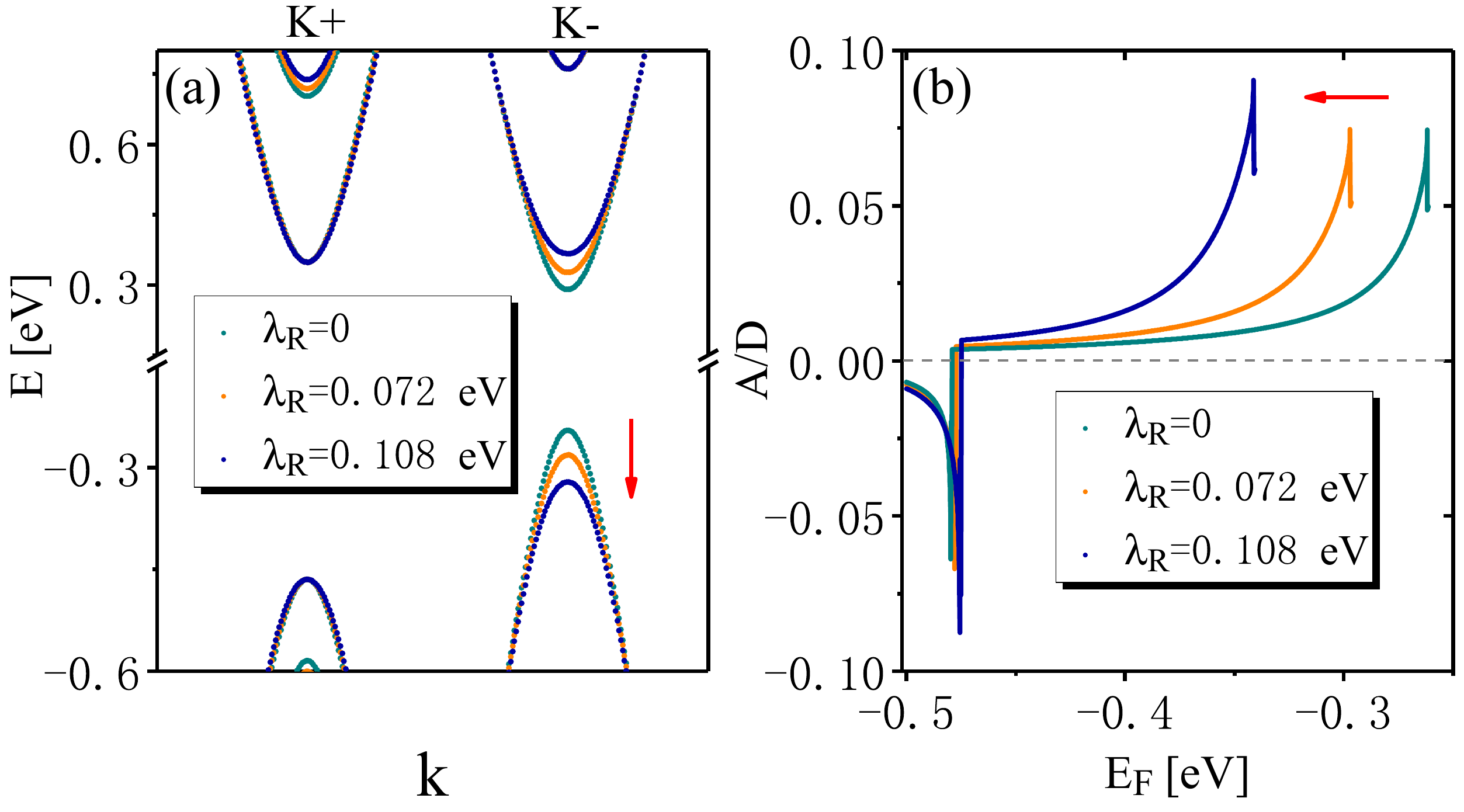}
\caption{ (a) Electronic band structure of monolayer MoTe$_2$ and (b) circular phonon dichroism $A/D$ versus the Fermi energy $E_F$ for different strength of Rashba spin-orbit coupling.}
\label{fig:cpd_rashba}
\end{figure}

$\textit{Roles of Rashba spin-orbit coupling}$.---Now we take into account the Rashba term $H_R$ and treat it as a perturbation. Analytical expressions are calculated~\cite{supple} and numerical results of electronic band structure and CPD are shown in Fig. \ref{fig:cpd_rashba}. In Fig. \ref{fig:cpd_rashba} (a), we find that the Rashba term shifts the conduction (valence) band edge to higher (lower) energy at valley $K_-$, whereas it hardly changes the band structure at valley $K_+$. This explains the phenomenon of peak shift in Fig. \ref{fig:cpd_rashba} (b) since the peaks are always close to the band edge. By increasing the strength of Rashba spin-orbit coupling, the magnitude of CPD can be enhanced, which provides us a knob to tune the CPD. For a realistic strength $\lambda_R=0.072$ eV~\cite{qi2015}, the behavior of $A/D$ is similar to the case without Rashba spin-orbit coupling, thus validating our above treatment. Particularly, we find that the introduction of $H_R$ does not change the reciprocal behaviors of absorption coefficients $\gamma^{L/R}$ under $\bm q\rightarrow-\bm q$. Therefore, to obtain the nonreciprocity, additional ingredients should be taken into account, such as the cyclotron motion of electrons~\cite{heil1982,kossacki2012,kumaravadivel2019,liu2020,sonntag2021} or phonon-magnon coupling~\cite{shanj2020}.

$\textit{Discussion and conclusion}$.---We have studied the circular phonon dichroism in magnetic two-dimensional materials, i.e., monolayer MoTe$_2$ in proximity to the EuO substrate. Large dichroism signal is obtained for the pseudogauge-type electron-phonon coupling, even without introducing the Landau levels or Rashba spin-orbit coupling. Such a signal is reciprocal (nonreciprocal) upon reversing the direction of phonon propagation (magnetization). By varying the gate voltage, the CPD signal can be tuned through the role of Fermi level and Rashba spin-orbit coupling. 

The proposed CPD effect can also be applied to other transition-metal dichalcogenides with spin-valley-coupled band structure, or their van der waals heterostructures. The effect can be detected by the pulse-echo technique~\cite{truell1969,luthi2004} based on the different absorption coefficients between left- and right-handed circularly polarized phonons. An alternative detection is the Raman spectroscopy analysis~\cite{kossacki2012,sonntag2021} of phonon polarization by injecting a linearly polarized acoustic waves. 

$\textit{Acknowledgments}$. This work is supported by the National Natural Science Foundation of China (NSFC, Grant No. 11904062). We also acknowledge the support of a startup grant from Guangzhou University.


\begin{thebibliography}{45}%
\makeatletter
\providecommand \@ifxundefined [1]{%
 \@ifx{#1\undefined}
}%
\providecommand \@ifnum [1]{%
 \ifnum #1\expandafter \@firstoftwo
 \else \expandafter \@secondoftwo
 \fi
}%
\providecommand \@ifx [1]{%
 \ifx #1\expandafter \@firstoftwo
 \else \expandafter \@secondoftwo
 \fi
}%
\providecommand \natexlab [1]{#1}%
\providecommand \enquote  [1]{``#1''}%
\providecommand \bibnamefont  [1]{#1}%
\providecommand \bibfnamefont [1]{#1}%
\providecommand \citenamefont [1]{#1}%
\providecommand \href@noop [0]{\@secondoftwo}%
\providecommand \href [0]{\begingroup \@sanitize@url \@href}%
\providecommand \@href[1]{\@@startlink{#1}\@@href}%
\providecommand \@@href[1]{\endgroup#1\@@endlink}%
\providecommand \@sanitize@url [0]{\catcode `\\12\catcode `\$12\catcode
  `\&12\catcode `\#12\catcode `\^12\catcode `\_12\catcode `\%12\relax}%
\providecommand \@@startlink[1]{}%
\providecommand \@@endlink[0]{}%
\providecommand \url  [0]{\begingroup\@sanitize@url \@url }%
\providecommand \@url [1]{\endgroup\@href {#1}{\urlprefix }}%
\providecommand \urlprefix  [0]{URL }%
\providecommand \Eprint [0]{\href }%
\providecommand \doibase [0]{http://dx.doi.org/}%
\providecommand \selectlanguage [0]{\@gobble}%
\providecommand \bibinfo  [0]{\@secondoftwo}%
\providecommand \bibfield  [0]{\@secondoftwo}%
\providecommand \translation [1]{[#1]}%
\providecommand \BibitemOpen [0]{}%
\providecommand \bibitemStop [0]{}%
\providecommand \bibitemNoStop [0]{.\EOS\space}%
\providecommand \EOS [0]{\spacefactor3000\relax}%
\providecommand \BibitemShut  [1]{\csname bibitem#1\endcsname}%
\let\auto@bib@innerbib\@empty
\bibitem [{\citenamefont {Zhang}\ and\ \citenamefont {Niu}(2015)}]{zhang2015}%
  \BibitemOpen
  \bibfield  {author} {\bibinfo {author} {\bibfnamefont {L.}~\bibnamefont
  {Zhang}}\ and\ \bibinfo {author} {\bibfnamefont {Q.}~\bibnamefont {Niu}},\
  }\href {\doibase 10.1103/PhysRevLett.115.115502} {\bibfield  {journal}
  {\bibinfo  {journal} {Phys. Rev. Lett.}\ }\textbf {\bibinfo {volume} {115}},\
  \bibinfo {pages} {115502} (\bibinfo {year} {2015})}\BibitemShut {NoStop}%
\bibitem [{\citenamefont {Zhu}\ \emph {et~al.}(2018)\citenamefont {Zhu},
  \citenamefont {Yi}, \citenamefont {Li}, \citenamefont {Xiao}, \citenamefont
  {Zhang}, \citenamefont {Yang}, \citenamefont {Kaindl}, \citenamefont {Li},
  \citenamefont {Wang},\ and\ \citenamefont {Zhang}}]{zhu2018}%
  \BibitemOpen
  \bibfield  {author} {\bibinfo {author} {\bibfnamefont {H.}~\bibnamefont
  {Zhu}}, \bibinfo {author} {\bibfnamefont {J.}~\bibnamefont {Yi}}, \bibinfo
  {author} {\bibfnamefont {M.-Y.}\ \bibnamefont {Li}}, \bibinfo {author}
  {\bibfnamefont {J.}~\bibnamefont {Xiao}}, \bibinfo {author} {\bibfnamefont
  {L.}~\bibnamefont {Zhang}}, \bibinfo {author} {\bibfnamefont {C.-W.}\
  \bibnamefont {Yang}}, \bibinfo {author} {\bibfnamefont {R.~A.}\ \bibnamefont
  {Kaindl}}, \bibinfo {author} {\bibfnamefont {L.-J.}\ \bibnamefont {Li}},
  \bibinfo {author} {\bibfnamefont {Y.}~\bibnamefont {Wang}}, \ and\ \bibinfo
  {author} {\bibfnamefont {X.}~\bibnamefont {Zhang}},\ }\href {\doibase
  10.1126/science.aar2711} {\bibfield  {journal} {\bibinfo  {journal}
  {Science}\ }\textbf {\bibinfo {volume} {359}},\ \bibinfo {pages} {579}
  (\bibinfo {year} {2018})}\BibitemShut {NoStop}%
\bibitem [{\citenamefont {Chen}\ \emph {et~al.}(2019)\citenamefont {Chen},
  \citenamefont {Lu}, \citenamefont {Dubey}, \citenamefont {Yao}, \citenamefont
  {Liu}, \citenamefont {Wang}, \citenamefont {Xiong}, \citenamefont {Zhang},\
  and\ \citenamefont {Srivastava}}]{chen2019}%
  \BibitemOpen
  \bibfield  {author} {\bibinfo {author} {\bibfnamefont {X.}~\bibnamefont
  {Chen}}, \bibinfo {author} {\bibfnamefont {X.}~\bibnamefont {Lu}}, \bibinfo
  {author} {\bibfnamefont {S.}~\bibnamefont {Dubey}}, \bibinfo {author}
  {\bibfnamefont {Q.}~\bibnamefont {Yao}}, \bibinfo {author} {\bibfnamefont
  {S.}~\bibnamefont {Liu}}, \bibinfo {author} {\bibfnamefont {X.}~\bibnamefont
  {Wang}}, \bibinfo {author} {\bibfnamefont {Q.}~\bibnamefont {Xiong}},
  \bibinfo {author} {\bibfnamefont {L.}~\bibnamefont {Zhang}}, \ and\ \bibinfo
  {author} {\bibfnamefont {A.}~\bibnamefont {Srivastava}},\ }\href {\doibase
  10.1038/s41567-018-0366-7} {\bibfield  {journal} {\bibinfo  {journal} {Nature
  Physics}\ }\textbf {\bibinfo {volume} {15}},\ \bibinfo {pages} {221}
  (\bibinfo {year} {2019})}\BibitemShut {NoStop}%
\bibitem [{\citenamefont {Maity}\ \emph {et~al.}()\citenamefont {Maity},
  \citenamefont {Mostofi},\ and\ \citenamefont {Lischner}}]{maity2021}%
  \BibitemOpen
  \bibfield  {author} {\bibinfo {author} {\bibfnamefont {I.}~\bibnamefont
  {Maity}}, \bibinfo {author} {\bibfnamefont {A.~A.}\ \bibnamefont {Mostofi}},
  \ and\ \bibinfo {author} {\bibfnamefont {J.}~\bibnamefont {Lischner}},\
  }\href@noop {} {\ }\Eprint {http://arxiv.org/abs/2108.03965}
  {arXiv:2108.03965} \BibitemShut {NoStop}%
\bibitem [{\citenamefont {Zhang}\ and\ \citenamefont {Niu}(2014)}]{zhang2014}%
  \BibitemOpen
  \bibfield  {author} {\bibinfo {author} {\bibfnamefont {L.}~\bibnamefont
  {Zhang}}\ and\ \bibinfo {author} {\bibfnamefont {Q.}~\bibnamefont {Niu}},\
  }\href {\doibase 10.1103/PhysRevLett.112.085503} {\bibfield  {journal}
  {\bibinfo  {journal} {Phys. Rev. Lett.}\ }\textbf {\bibinfo {volume} {112}},\
  \bibinfo {pages} {085503} (\bibinfo {year} {2014})}\BibitemShut {NoStop}%
\bibitem [{\citenamefont {Hamada}\ \emph {et~al.}(2018)\citenamefont {Hamada},
  \citenamefont {Minamitani}, \citenamefont {Hirayama},\ and\ \citenamefont
  {Murakami}}]{hamada2018}%
  \BibitemOpen
  \bibfield  {author} {\bibinfo {author} {\bibfnamefont {M.}~\bibnamefont
  {Hamada}}, \bibinfo {author} {\bibfnamefont {E.}~\bibnamefont {Minamitani}},
  \bibinfo {author} {\bibfnamefont {M.}~\bibnamefont {Hirayama}}, \ and\
  \bibinfo {author} {\bibfnamefont {S.}~\bibnamefont {Murakami}},\ }\href
  {\doibase 10.1103/PhysRevLett.121.175301} {\bibfield  {journal} {\bibinfo
  {journal} {Phys. Rev. Lett.}\ }\textbf {\bibinfo {volume} {121}},\ \bibinfo
  {pages} {175301} (\bibinfo {year} {2018})}\BibitemShut {NoStop}%
\bibitem [{\citenamefont {Zhang}\ and\ \citenamefont {Murakami}()}]{zhang2021}%
  \BibitemOpen
  \bibfield  {author} {\bibinfo {author} {\bibfnamefont {T.}~\bibnamefont
  {Zhang}}\ and\ \bibinfo {author} {\bibfnamefont {S.}~\bibnamefont
  {Murakami}},\ }\href@noop {} {\ }\Eprint {http://arxiv.org/abs/2107.04243}
  {arXiv:2107.04243} \BibitemShut {NoStop}%
\bibitem [{\citenamefont {Komiyama}\ and\ \citenamefont
  {Murakami}(2021)}]{komiyama2021}%
  \BibitemOpen
  \bibfield  {author} {\bibinfo {author} {\bibfnamefont {H.}~\bibnamefont
  {Komiyama}}\ and\ \bibinfo {author} {\bibfnamefont {S.}~\bibnamefont
  {Murakami}},\ }\href {\doibase 10.1103/PhysRevB.103.214302} {\bibfield
  {journal} {\bibinfo  {journal} {Phys. Rev. B}\ }\textbf {\bibinfo {volume}
  {103}},\ \bibinfo {pages} {214302} (\bibinfo {year} {2021})}\BibitemShut
  {NoStop}%
\bibitem [{\citenamefont {Juraschek}\ \emph {et~al.}(2017)\citenamefont
  {Juraschek}, \citenamefont {Fechner}, \citenamefont {Balatsky},\ and\
  \citenamefont {Spaldin}}]{juraschek2017}%
  \BibitemOpen
  \bibfield  {author} {\bibinfo {author} {\bibfnamefont {D.~M.}\ \bibnamefont
  {Juraschek}}, \bibinfo {author} {\bibfnamefont {M.}~\bibnamefont {Fechner}},
  \bibinfo {author} {\bibfnamefont {A.~V.}\ \bibnamefont {Balatsky}}, \ and\
  \bibinfo {author} {\bibfnamefont {N.~A.}\ \bibnamefont {Spaldin}},\ }\href
  {\doibase 10.1103/PhysRevMaterials.1.014401} {\bibfield  {journal} {\bibinfo
  {journal} {Phys. Rev. Materials}\ }\textbf {\bibinfo {volume} {1}},\ \bibinfo
  {pages} {014401} (\bibinfo {year} {2017})}\BibitemShut {NoStop}%
\bibitem [{\citenamefont {Juraschek}\ and\ \citenamefont
  {Spaldin}(2019)}]{juraschek2019}%
  \BibitemOpen
  \bibfield  {author} {\bibinfo {author} {\bibfnamefont {D.~M.}\ \bibnamefont
  {Juraschek}}\ and\ \bibinfo {author} {\bibfnamefont {N.~A.}\ \bibnamefont
  {Spaldin}},\ }\href {\doibase 10.1103/PhysRevMaterials.3.064405} {\bibfield
  {journal} {\bibinfo  {journal} {Phys. Rev. Materials}\ }\textbf {\bibinfo
  {volume} {3}},\ \bibinfo {pages} {064405} (\bibinfo {year}
  {2019})}\BibitemShut {NoStop}%
\bibitem [{\citenamefont {Cheng}\ \emph {et~al.}(2020)\citenamefont {Cheng},
  \citenamefont {Schumann}, \citenamefont {Wang}, \citenamefont {Zhang},
  \citenamefont {Barbalas}, \citenamefont {Stemmer},\ and\ \citenamefont
  {Armitage}}]{cheng2020}%
  \BibitemOpen
  \bibfield  {author} {\bibinfo {author} {\bibfnamefont {B.}~\bibnamefont
  {Cheng}}, \bibinfo {author} {\bibfnamefont {T.}~\bibnamefont {Schumann}},
  \bibinfo {author} {\bibfnamefont {Y.~C.}\ \bibnamefont {Wang}}, \bibinfo
  {author} {\bibfnamefont {X.~S.}\ \bibnamefont {Zhang}}, \bibinfo {author}
  {\bibfnamefont {D.}~\bibnamefont {Barbalas}}, \bibinfo {author}
  {\bibfnamefont {S.}~\bibnamefont {Stemmer}}, \ and\ \bibinfo {author}
  {\bibfnamefont {N.~P.}\ \bibnamefont {Armitage}},\ }\href {\doibase
  10.1021/acs.nanolett.0c01983} {\bibfield  {journal} {\bibinfo  {journal}
  {Nano Lett.}\ }\textbf {\bibinfo {volume} {20}},\ \bibinfo {pages} {5991}
  (\bibinfo {year} {2020})}\BibitemShut {NoStop}%
\bibitem [{\citenamefont {Juraschek}\ \emph {et~al.}(2020)\citenamefont
  {Juraschek}, \citenamefont {Narang},\ and\ \citenamefont
  {Spaldin}}]{juraschek2020}%
  \BibitemOpen
  \bibfield  {author} {\bibinfo {author} {\bibfnamefont {D.~M.}\ \bibnamefont
  {Juraschek}}, \bibinfo {author} {\bibfnamefont {P.}~\bibnamefont {Narang}}, \
  and\ \bibinfo {author} {\bibfnamefont {N.~A.}\ \bibnamefont {Spaldin}},\
  }\href {\doibase 10.1103/PhysRevResearch.2.043035} {\bibfield  {journal}
  {\bibinfo  {journal} {Phys. Rev. Research}\ }\textbf {\bibinfo {volume}
  {2}},\ \bibinfo {pages} {043035} (\bibinfo {year} {2020})}\BibitemShut
  {NoStop}%
\bibitem [{\citenamefont {Park}\ and\ \citenamefont {Yang}(2020)}]{park2020}%
  \BibitemOpen
  \bibfield  {author} {\bibinfo {author} {\bibfnamefont {S.}~\bibnamefont
  {Park}}\ and\ \bibinfo {author} {\bibfnamefont {B.-J.}\ \bibnamefont
  {Yang}},\ }\href {\doibase 10.1021/acs.nanolett.0c03220} {\bibfield
  {journal} {\bibinfo  {journal} {Nano Lett.}\ }\textbf {\bibinfo {volume}
  {20}},\ \bibinfo {pages} {7694} (\bibinfo {year} {2020})}\BibitemShut
  {NoStop}%
\bibitem [{\citenamefont {Hamada}\ and\ \citenamefont
  {Murakami}(2020)}]{hamada2020}%
  \BibitemOpen
  \bibfield  {author} {\bibinfo {author} {\bibfnamefont {M.}~\bibnamefont
  {Hamada}}\ and\ \bibinfo {author} {\bibfnamefont {S.}~\bibnamefont
  {Murakami}},\ }\href {\doibase 10.1103/PhysRevB.101.144306} {\bibfield
  {journal} {\bibinfo  {journal} {Phys. Rev. B}\ }\textbf {\bibinfo {volume}
  {101}},\ \bibinfo {pages} {144306} (\bibinfo {year} {2020})}\BibitemShut
  {NoStop}%
\bibitem [{\citenamefont {Hu}\ \emph {et~al.}(2021)\citenamefont {Hu},
  \citenamefont {Yu}, \citenamefont {Garate},\ and\ \citenamefont
  {Liu}}]{hu2021}%
  \BibitemOpen
  \bibfield  {author} {\bibinfo {author} {\bibfnamefont {L.-H.}\ \bibnamefont
  {Hu}}, \bibinfo {author} {\bibfnamefont {J.}~\bibnamefont {Yu}}, \bibinfo
  {author} {\bibfnamefont {I.}~\bibnamefont {Garate}}, \ and\ \bibinfo {author}
  {\bibfnamefont {C.-X.}\ \bibnamefont {Liu}},\ }\href {\doibase
  10.1103/PhysRevLett.127.125901} {\bibfield  {journal} {\bibinfo  {journal}
  {Phys. Rev. Lett.}\ }\textbf {\bibinfo {volume} {127}},\ \bibinfo {pages}
  {125901} (\bibinfo {year} {2021})}\BibitemShut {NoStop}%
\bibitem [{\citenamefont {Nomura}\ \emph {et~al.}(2019)\citenamefont {Nomura},
  \citenamefont {Zhang}, \citenamefont {Zherlitsyn}, \citenamefont {Wosnitza},
  \citenamefont {Tokura}, \citenamefont {Nagaosa},\ and\ \citenamefont
  {Seki}}]{nomura2019}%
  \BibitemOpen
  \bibfield  {author} {\bibinfo {author} {\bibfnamefont {T.}~\bibnamefont
  {Nomura}}, \bibinfo {author} {\bibfnamefont {X.-X.}\ \bibnamefont {Zhang}},
  \bibinfo {author} {\bibfnamefont {S.}~\bibnamefont {Zherlitsyn}}, \bibinfo
  {author} {\bibfnamefont {J.}~\bibnamefont {Wosnitza}}, \bibinfo {author}
  {\bibfnamefont {Y.}~\bibnamefont {Tokura}}, \bibinfo {author} {\bibfnamefont
  {N.}~\bibnamefont {Nagaosa}}, \ and\ \bibinfo {author} {\bibfnamefont
  {S.}~\bibnamefont {Seki}},\ }\href {\doibase 10.1103/PhysRevLett.122.145901}
  {\bibfield  {journal} {\bibinfo  {journal} {Phys. Rev. Lett.}\ }\textbf
  {\bibinfo {volume} {122}},\ \bibinfo {pages} {145901} (\bibinfo {year}
  {2019})}\BibitemShut {NoStop}%
\bibitem [{\citenamefont {Sengupta}\ \emph {et~al.}(2020)\citenamefont
  {Sengupta}, \citenamefont {Lhachemi},\ and\ \citenamefont
  {Garate}}]{sengupta2020}%
  \BibitemOpen
  \bibfield  {author} {\bibinfo {author} {\bibfnamefont {S.}~\bibnamefont
  {Sengupta}}, \bibinfo {author} {\bibfnamefont {M.~N.~Y.}\ \bibnamefont
  {Lhachemi}}, \ and\ \bibinfo {author} {\bibfnamefont {I.}~\bibnamefont
  {Garate}},\ }\href {\doibase 10.1103/PhysRevLett.125.146402} {\bibfield
  {journal} {\bibinfo  {journal} {Phys. Rev. Lett.}\ }\textbf {\bibinfo
  {volume} {125}},\ \bibinfo {pages} {146402} (\bibinfo {year}
  {2020})}\BibitemShut {NoStop}%
\bibitem [{\citenamefont {Souza}\ and\ \citenamefont
  {Vanderbilt}(2008)}]{souza2008}%
  \BibitemOpen
  \bibfield  {author} {\bibinfo {author} {\bibfnamefont {I.}~\bibnamefont
  {Souza}}\ and\ \bibinfo {author} {\bibfnamefont {D.}~\bibnamefont
  {Vanderbilt}},\ }\href {\doibase 10.1103/PhysRevB.77.054438} {\bibfield
  {journal} {\bibinfo  {journal} {Phys. Rev. B}\ }\textbf {\bibinfo {volume}
  {77}},\ \bibinfo {pages} {054438} (\bibinfo {year} {2008})}\BibitemShut
  {NoStop}%
\bibitem [{\citenamefont {Wang}\ \emph {et~al.}(2011)\citenamefont {Wang},
  \citenamefont {Hsieh}, \citenamefont {Pilon}, \citenamefont {Fu},
  \citenamefont {Gardner}, \citenamefont {Lee},\ and\ \citenamefont
  {Gedik}}]{wang2011}%
  \BibitemOpen
  \bibfield  {author} {\bibinfo {author} {\bibfnamefont {Y.~H.}\ \bibnamefont
  {Wang}}, \bibinfo {author} {\bibfnamefont {D.}~\bibnamefont {Hsieh}},
  \bibinfo {author} {\bibfnamefont {D.}~\bibnamefont {Pilon}}, \bibinfo
  {author} {\bibfnamefont {L.}~\bibnamefont {Fu}}, \bibinfo {author}
  {\bibfnamefont {D.~R.}\ \bibnamefont {Gardner}}, \bibinfo {author}
  {\bibfnamefont {Y.~S.}\ \bibnamefont {Lee}}, \ and\ \bibinfo {author}
  {\bibfnamefont {N.}~\bibnamefont {Gedik}},\ }\href {\doibase
  10.1103/PhysRevLett.107.207602} {\bibfield  {journal} {\bibinfo  {journal}
  {Phys. Rev. Lett.}\ }\textbf {\bibinfo {volume} {107}},\ \bibinfo {pages}
  {207602} (\bibinfo {year} {2011})}\BibitemShut {NoStop}%
\bibitem [{\citenamefont {Tran}\ \emph {et~al.}(2017)\citenamefont {Tran},
  \citenamefont {Dauphin}, \citenamefont {Grushin}, \citenamefont {Zoller},\
  and\ \citenamefont {Goldman}}]{tran2017}%
  \BibitemOpen
  \bibfield  {author} {\bibinfo {author} {\bibfnamefont {D.~T.}\ \bibnamefont
  {Tran}}, \bibinfo {author} {\bibfnamefont {A.}~\bibnamefont {Dauphin}},
  \bibinfo {author} {\bibfnamefont {A.~G.}\ \bibnamefont {Grushin}}, \bibinfo
  {author} {\bibfnamefont {P.}~\bibnamefont {Zoller}}, \ and\ \bibinfo {author}
  {\bibfnamefont {N.}~\bibnamefont {Goldman}},\ }\href {\doibase
  10.1126/sciadv.1701207} {\bibfield  {journal} {\bibinfo  {journal} {Sci.
  Adv.}\ }\textbf {\bibinfo {volume} {3}},\ \bibinfo {pages} {e1701207}
  (\bibinfo {year} {2017})}\BibitemShut {NoStop}%
\bibitem [{\citenamefont {Liu}\ \emph {et~al.}(2018)\citenamefont {Liu},
  \citenamefont {Yang},\ and\ \citenamefont {Zhang}}]{liu2018}%
  \BibitemOpen
  \bibfield  {author} {\bibinfo {author} {\bibfnamefont {Y.}~\bibnamefont
  {Liu}}, \bibinfo {author} {\bibfnamefont {S.~A.}\ \bibnamefont {Yang}}, \
  and\ \bibinfo {author} {\bibfnamefont {F.}~\bibnamefont {Zhang}},\ }\href
  {\doibase 10.1103/PhysRevB.97.035153} {\bibfield  {journal} {\bibinfo
  {journal} {Phys. Rev. B}\ }\textbf {\bibinfo {volume} {97}},\ \bibinfo
  {pages} {035153} (\bibinfo {year} {2018})}\BibitemShut {NoStop}%
\bibitem [{\citenamefont {Pozo}\ \emph {et~al.}(2019)\citenamefont {Pozo},
  \citenamefont {Repellin},\ and\ \citenamefont {Grushin}}]{pozo2019}%
  \BibitemOpen
  \bibfield  {author} {\bibinfo {author} {\bibfnamefont {O.}~\bibnamefont
  {Pozo}}, \bibinfo {author} {\bibfnamefont {C.}~\bibnamefont {Repellin}}, \
  and\ \bibinfo {author} {\bibfnamefont {A.~G.}\ \bibnamefont {Grushin}},\
  }\href {\doibase 10.1103/PhysRevLett.123.247401} {\bibfield  {journal}
  {\bibinfo  {journal} {Phys. Rev. Lett.}\ }\textbf {\bibinfo {volume} {123}},\
  \bibinfo {pages} {247401} (\bibinfo {year} {2019})}\BibitemShut {NoStop}%
\bibitem [{\citenamefont {Repellin}\ and\ \citenamefont
  {Goldman}(2019)}]{repellin2019}%
  \BibitemOpen
  \bibfield  {author} {\bibinfo {author} {\bibfnamefont {C.}~\bibnamefont
  {Repellin}}\ and\ \bibinfo {author} {\bibfnamefont {N.}~\bibnamefont
  {Goldman}},\ }\href {\doibase 10.1103/PhysRevLett.122.166801} {\bibfield
  {journal} {\bibinfo  {journal} {Phys. Rev. Lett.}\ }\textbf {\bibinfo
  {volume} {122}},\ \bibinfo {pages} {166801} (\bibinfo {year}
  {2019})}\BibitemShut {NoStop}%
\bibitem [{\citenamefont {Sch\"{u}ler}\ \emph {et~al.}(2020)\citenamefont
  {Sch\"{u}ler}, \citenamefont {Giovannini}, \citenamefont {H\"{u}bener},
  \citenamefont {Rubio}, \citenamefont {Sentef},\ and\ \citenamefont
  {Werner}}]{schuler2020}%
  \BibitemOpen
  \bibfield  {author} {\bibinfo {author} {\bibfnamefont {M.}~\bibnamefont
  {Sch\"{u}ler}}, \bibinfo {author} {\bibfnamefont {U.~D.}\ \bibnamefont
  {Giovannini}}, \bibinfo {author} {\bibfnamefont {H.}~\bibnamefont
  {H\"{u}bener}}, \bibinfo {author} {\bibfnamefont {A.}~\bibnamefont {Rubio}},
  \bibinfo {author} {\bibfnamefont {M.~A.}\ \bibnamefont {Sentef}}, \ and\
  \bibinfo {author} {\bibfnamefont {P.}~\bibnamefont {Werner}},\ }\href
  {\doibase 10.1126/sciadv.aay2730} {\bibfield  {journal} {\bibinfo  {journal}
  {Sci. Adv.}\ }\textbf {\bibinfo {volume} {6}},\ \bibinfo {pages} {eaay2730}
  (\bibinfo {year} {2020})}\BibitemShut {NoStop}%
\bibitem [{\citenamefont {Liu}\ and\ \citenamefont {Shi}(2017)}]{liu2017}%
  \BibitemOpen
  \bibfield  {author} {\bibinfo {author} {\bibfnamefont {D.}~\bibnamefont
  {Liu}}\ and\ \bibinfo {author} {\bibfnamefont {J.}~\bibnamefont {Shi}},\
  }\href {\doibase 10.1103/PhysRevLett.119.075301} {\bibfield  {journal}
  {\bibinfo  {journal} {Phys. Rev. Lett.}\ }\textbf {\bibinfo {volume} {119}},\
  \bibinfo {pages} {075301} (\bibinfo {year} {2017})}\BibitemShut {NoStop}%
\bibitem [{\citenamefont {Sonntag}\ \emph {et~al.}(2021)\citenamefont
  {Sonntag}, \citenamefont {Reichardt}, \citenamefont {Beschoten},\ and\
  \citenamefont {Stampfer}}]{sonntag2021}%
  \BibitemOpen
  \bibfield  {author} {\bibinfo {author} {\bibfnamefont {J.}~\bibnamefont
  {Sonntag}}, \bibinfo {author} {\bibfnamefont {S.}~\bibnamefont {Reichardt}},
  \bibinfo {author} {\bibfnamefont {B.}~\bibnamefont {Beschoten}}, \ and\
  \bibinfo {author} {\bibfnamefont {C.}~\bibnamefont {Stampfer}},\ }\href
  {\doibase 10.1021/acs.nanolett.0c05043} {\bibfield  {journal} {\bibinfo
  {journal} {Nano Lett.}\ }\textbf {\bibinfo {volume} {21}},\ \bibinfo {pages}
  {2898} (\bibinfo {year} {2021})}\BibitemShut {NoStop}%
\bibitem [{\citenamefont {Kossacki}\ \emph {et~al.}(2012)\citenamefont
  {Kossacki}, \citenamefont {Faugeras}, \citenamefont {K\"uhne}, \citenamefont
  {Orlita}, \citenamefont {Mahmood}, \citenamefont {Dujardin}, \citenamefont
  {Nair}, \citenamefont {Geim},\ and\ \citenamefont {Potemski}}]{kossacki2012}%
  \BibitemOpen
  \bibfield  {author} {\bibinfo {author} {\bibfnamefont {P.}~\bibnamefont
  {Kossacki}}, \bibinfo {author} {\bibfnamefont {C.}~\bibnamefont {Faugeras}},
  \bibinfo {author} {\bibfnamefont {M.}~\bibnamefont {K\"uhne}}, \bibinfo
  {author} {\bibfnamefont {M.}~\bibnamefont {Orlita}}, \bibinfo {author}
  {\bibfnamefont {A.}~\bibnamefont {Mahmood}}, \bibinfo {author} {\bibfnamefont
  {E.}~\bibnamefont {Dujardin}}, \bibinfo {author} {\bibfnamefont {R.~R.}\
  \bibnamefont {Nair}}, \bibinfo {author} {\bibfnamefont {A.~K.}\ \bibnamefont
  {Geim}}, \ and\ \bibinfo {author} {\bibfnamefont {M.}~\bibnamefont
  {Potemski}},\ }\href {\doibase 10.1103/PhysRevB.86.205431} {\bibfield
  {journal} {\bibinfo  {journal} {Phys. Rev. B}\ }\textbf {\bibinfo {volume}
  {86}},\ \bibinfo {pages} {205431} (\bibinfo {year} {2012})}\BibitemShut
  {NoStop}%
\bibitem [{\citenamefont {Kumaravadivel}\ \emph {et~al.}(2019)\citenamefont
  {Kumaravadivel}, \citenamefont {Greenaway}, \citenamefont {Perello},
  \citenamefont {Berdyugin}, \citenamefont {J.}, \citenamefont {Wengraf},
  \citenamefont {Liu}, \citenamefont {Edgar}, \citenamefont {Geim},
  \citenamefont {Eaves},\ and\ \citenamefont {Kumar}}]{kumaravadivel2019}%
  \BibitemOpen
  \bibfield  {author} {\bibinfo {author} {\bibfnamefont {P.}~\bibnamefont
  {Kumaravadivel}}, \bibinfo {author} {\bibfnamefont {M.~T.}\ \bibnamefont
  {Greenaway}}, \bibinfo {author} {\bibfnamefont {D.}~\bibnamefont {Perello}},
  \bibinfo {author} {\bibfnamefont {A.}~\bibnamefont {Berdyugin}}, \bibinfo
  {author} {\bibfnamefont {B.}~\bibnamefont {J.}}, \bibinfo {author}
  {\bibfnamefont {J.}~\bibnamefont {Wengraf}}, \bibinfo {author} {\bibfnamefont
  {S.}~\bibnamefont {Liu}}, \bibinfo {author} {\bibfnamefont {J.~H.}\
  \bibnamefont {Edgar}}, \bibinfo {author} {\bibfnamefont {A.~K.}\ \bibnamefont
  {Geim}}, \bibinfo {author} {\bibfnamefont {L.}~\bibnamefont {Eaves}}, \ and\
  \bibinfo {author} {\bibfnamefont {R.~K.}\ \bibnamefont {Kumar}},\ }\href
  {\doibase 10.1038/s41467-019-11379-3} {\bibfield  {journal} {\bibinfo
  {journal} {Nat. Commun.}\ }\textbf {\bibinfo {volume} {10}},\ \bibinfo
  {pages} {3334} (\bibinfo {year} {2019})}\BibitemShut {NoStop}%
\bibitem [{\citenamefont {Qi}\ \emph {et~al.}(2015)\citenamefont {Qi},
  \citenamefont {Li}, \citenamefont {Niu},\ and\ \citenamefont
  {Feng}}]{qi2015}%
  \BibitemOpen
  \bibfield  {author} {\bibinfo {author} {\bibfnamefont {J.}~\bibnamefont
  {Qi}}, \bibinfo {author} {\bibfnamefont {X.}~\bibnamefont {Li}}, \bibinfo
  {author} {\bibfnamefont {Q.}~\bibnamefont {Niu}}, \ and\ \bibinfo {author}
  {\bibfnamefont {J.}~\bibnamefont {Feng}},\ }\href {\doibase
  10.1103/PhysRevB.92.121403} {\bibfield  {journal} {\bibinfo  {journal} {Phys.
  Rev. B}\ }\textbf {\bibinfo {volume} {92}},\ \bibinfo {pages} {121403}
  (\bibinfo {year} {2015})}\BibitemShut {NoStop}%
\bibitem [{\citenamefont {Habe}\ and\ \citenamefont
  {Koshino}(2017)}]{habe2017}%
  \BibitemOpen
  \bibfield  {author} {\bibinfo {author} {\bibfnamefont {T.}~\bibnamefont
  {Habe}}\ and\ \bibinfo {author} {\bibfnamefont {M.}~\bibnamefont {Koshino}},\
  }\href {\doibase 10.1103/PhysRevB.96.085411} {\bibfield  {journal} {\bibinfo
  {journal} {Phys. Rev. B}\ }\textbf {\bibinfo {volume} {96}},\ \bibinfo
  {pages} {085411} (\bibinfo {year} {2017})}\BibitemShut {NoStop}%
\bibitem [{lan()}]{landau1959}%
  \BibitemOpen
  \href@noop {} {}\bibinfo {note} {L. D. Landau and E. M. Lifschitz,
  \emph{Theory of Elasticity}, Pergamon Press, Oxford, 1959.}\BibitemShut
  {Stop}%
\bibitem [{\citenamefont {Suzuura}\ and\ \citenamefont
  {Ando}(2002)}]{suzuura2002}%
  \BibitemOpen
  \bibfield  {author} {\bibinfo {author} {\bibfnamefont {H.}~\bibnamefont
  {Suzuura}}\ and\ \bibinfo {author} {\bibfnamefont {T.}~\bibnamefont {Ando}},\
  }\href {\doibase 10.1103/PhysRevB.65.235412} {\bibfield  {journal} {\bibinfo
  {journal} {Phys. Rev. B}\ }\textbf {\bibinfo {volume} {65}},\ \bibinfo
  {pages} {235412} (\bibinfo {year} {2002})}\BibitemShut {NoStop}%
\bibitem [{\citenamefont {Cazalilla}\ \emph {et~al.}(2014)\citenamefont
  {Cazalilla}, \citenamefont {Ochoa},\ and\ \citenamefont
  {Guinea}}]{cazalilla2014}%
  \BibitemOpen
  \bibfield  {author} {\bibinfo {author} {\bibfnamefont {M.~A.}\ \bibnamefont
  {Cazalilla}}, \bibinfo {author} {\bibfnamefont {H.}~\bibnamefont {Ochoa}}, \
  and\ \bibinfo {author} {\bibfnamefont {F.}~\bibnamefont {Guinea}},\ }\href
  {\doibase 10.1103/PhysRevLett.113.077201} {\bibfield  {journal} {\bibinfo
  {journal} {Phys. Rev. Lett.}\ }\textbf {\bibinfo {volume} {113}},\ \bibinfo
  {pages} {077201} (\bibinfo {year} {2014})}\BibitemShut {NoStop}%
\bibitem [{\citenamefont {Shan}(2020)}]{shan2020}%
  \BibitemOpen
  \bibfield  {author} {\bibinfo {author} {\bibfnamefont {W.-Y.}\ \bibnamefont
  {Shan}},\ }\href {\doibase 10.1103/PhysRevB.102.241301} {\bibfield  {journal}
  {\bibinfo  {journal} {Phys. Rev. B}\ }\textbf {\bibinfo {volume} {102}},\
  \bibinfo {pages} {241301} (\bibinfo {year} {2020})}\BibitemShut {NoStop}%
\bibitem [{sup()}]{supple}%
  \BibitemOpen
  \href@noop {} {}\bibinfo {note} {See Supplementary Material for calculation
  details.}\BibitemShut {Stop}%
\bibitem [{\citenamefont {Rano}\ \emph {et~al.}(2020)\citenamefont {Rano},
  \citenamefont {Syed},\ and\ \citenamefont {Naqib}}]{rano2020}%
  \BibitemOpen
  \bibfield  {author} {\bibinfo {author} {\bibfnamefont {B.~R.}\ \bibnamefont
  {Rano}}, \bibinfo {author} {\bibfnamefont {I.~M.}\ \bibnamefont {Syed}}, \
  and\ \bibinfo {author} {\bibfnamefont {S.~H.}\ \bibnamefont {Naqib}},\ }\href
  {\doibase 10.1016/j.rinp.2020.103639} {\bibfield  {journal} {\bibinfo
  {journal} {Results in Physics}\ }\textbf {\bibinfo {volume} {19}},\ \bibinfo
  {pages} {103639} (\bibinfo {year} {2020})}\BibitemShut {NoStop}%
\bibitem [{\citenamefont {Keum}\ \emph {et~al.}(2015)\citenamefont {Keum},
  \citenamefont {Cho}, \citenamefont {Kim}, \citenamefont {Choe}, \citenamefont
  {Sung}, \citenamefont {Kan}, \citenamefont {Kang}, \citenamefont {Hwang},
  \citenamefont {Kim}, \citenamefont {Yang}, \citenamefont {Chang},\ and\
  \citenamefont {Lee}}]{keum2015}%
  \BibitemOpen
  \bibfield  {author} {\bibinfo {author} {\bibfnamefont {D.~H.}\ \bibnamefont
  {Keum}}, \bibinfo {author} {\bibfnamefont {S.}~\bibnamefont {Cho}}, \bibinfo
  {author} {\bibfnamefont {J.~H.}\ \bibnamefont {Kim}}, \bibinfo {author}
  {\bibfnamefont {D.-H.}\ \bibnamefont {Choe}}, \bibinfo {author}
  {\bibfnamefont {H.-J.}\ \bibnamefont {Sung}}, \bibinfo {author}
  {\bibfnamefont {M.}~\bibnamefont {Kan}}, \bibinfo {author} {\bibfnamefont
  {H.}~\bibnamefont {Kang}}, \bibinfo {author} {\bibfnamefont {J.-Y.}\
  \bibnamefont {Hwang}}, \bibinfo {author} {\bibfnamefont {S.~W.}\ \bibnamefont
  {Kim}}, \bibinfo {author} {\bibfnamefont {H.}~\bibnamefont {Yang}}, \bibinfo
  {author} {\bibfnamefont {K.~J.}\ \bibnamefont {Chang}}, \ and\ \bibinfo
  {author} {\bibfnamefont {Y.~H.}\ \bibnamefont {Lee}},\ }\href {\doibase
  doi.org/10.1038/nphys3314} {\bibfield  {journal} {\bibinfo  {journal} {Nature
  Physics}\ }\textbf {\bibinfo {volume} {11}},\ \bibinfo {pages} {482}
  (\bibinfo {year} {2015})}\BibitemShut {NoStop}%
\bibitem [{\citenamefont {Chen}\ \emph {et~al.}(2016)\citenamefont {Chen},
  \citenamefont {Luo}, \citenamefont {Xiao}, \citenamefont {Lu}, \citenamefont
  {Zhang}, \citenamefont {Yang}, \citenamefont {Li}, \citenamefont {Pei},
  \citenamefont {Shao}, \citenamefont {Zhang}, \citenamefont {Ling},
  \citenamefont {Xi}, \citenamefont {Song},\ and\ \citenamefont
  {Sun}}]{chen2016}%
  \BibitemOpen
  \bibfield  {author} {\bibinfo {author} {\bibfnamefont {F.~C.}\ \bibnamefont
  {Chen}}, \bibinfo {author} {\bibfnamefont {X.}~\bibnamefont {Luo}}, \bibinfo
  {author} {\bibfnamefont {R.~C.}\ \bibnamefont {Xiao}}, \bibinfo {author}
  {\bibfnamefont {W.~J.}\ \bibnamefont {Lu}}, \bibinfo {author} {\bibfnamefont
  {B.}~\bibnamefont {Zhang}}, \bibinfo {author} {\bibfnamefont {H.~X.}\
  \bibnamefont {Yang}}, \bibinfo {author} {\bibfnamefont {J.~Q.}\ \bibnamefont
  {Li}}, \bibinfo {author} {\bibfnamefont {Q.~L.}\ \bibnamefont {Pei}},
  \bibinfo {author} {\bibfnamefont {D.~F.}\ \bibnamefont {Shao}}, \bibinfo
  {author} {\bibfnamefont {R.~R.}\ \bibnamefont {Zhang}}, \bibinfo {author}
  {\bibfnamefont {L.~S.}\ \bibnamefont {Ling}}, \bibinfo {author}
  {\bibfnamefont {C.~Y.}\ \bibnamefont {Xi}}, \bibinfo {author} {\bibfnamefont
  {W.~H.}\ \bibnamefont {Song}}, \ and\ \bibinfo {author} {\bibfnamefont
  {Y.~P.}\ \bibnamefont {Sun}},\ }\href {\doibase 10.1063/1.4947433} {\bibfield
   {journal} {\bibinfo  {journal} {Appl. Phys. Lett.}\ }\textbf {\bibinfo
  {volume} {108}},\ \bibinfo {pages} {162601} (\bibinfo {year}
  {2016})}\BibitemShut {NoStop}%
\bibitem [{\citenamefont {Freiser}(1968)}]{freiser19682002}%
  \BibitemOpen
  \bibfield  {author} {\bibinfo {author} {\bibfnamefont {M.}~\bibnamefont
  {Freiser}},\ }\href {\doibase 10.1109/TMAG.1968.1066210} {\bibfield
  {journal} {\bibinfo  {journal} {IEEE Trans. Magn.}\ }\textbf {\bibinfo
  {volume} {4}},\ \bibinfo {pages} {152} (\bibinfo {year} {1968})}\BibitemShut
  {NoStop}%
\bibitem [{\citenamefont {Fuchs}(1965)}]{fuchs1965}%
  \BibitemOpen
  \bibfield  {author} {\bibinfo {author} {\bibfnamefont {R.}~\bibnamefont
  {Fuchs}},\ }\href {\doibase 10.1080/14786436508224252} {\bibfield  {journal}
  {\bibinfo  {journal} {Phil. Mag.}\ }\textbf {\bibinfo {volume} {11}},\
  \bibinfo {pages} {647} (\bibinfo {year} {1965})}\BibitemShut {NoStop}%
\bibitem [{\citenamefont {Heil}\ \emph {et~al.}(1982)\citenamefont {Heil},
  \citenamefont {L\"uthi},\ and\ \citenamefont {Thalmeier}}]{heil1982}%
  \BibitemOpen
  \bibfield  {author} {\bibinfo {author} {\bibfnamefont {J.}~\bibnamefont
  {Heil}}, \bibinfo {author} {\bibfnamefont {B.}~\bibnamefont {L\"uthi}}, \
  and\ \bibinfo {author} {\bibfnamefont {P.}~\bibnamefont {Thalmeier}},\ }\href
  {\doibase 10.1103/PhysRevB.25.6515} {\bibfield  {journal} {\bibinfo
  {journal} {Phys. Rev. B}\ }\textbf {\bibinfo {volume} {25}},\ \bibinfo
  {pages} {6515} (\bibinfo {year} {1982})}\BibitemShut {NoStop}%
\bibitem [{\citenamefont {Liu}\ \emph {et~al.}(2020)\citenamefont {Liu},
  \citenamefont {Guo}, \citenamefont {Hu}, \citenamefont {Shi}, \citenamefont
  {Li}, \citenamefont {Zhang}, \citenamefont {Chen}, \citenamefont {Zhang},
  \citenamefont {Zhou}, \citenamefont {Lu}, \citenamefont {Lin}, \citenamefont
  {Liu}, \citenamefont {Cheng}, \citenamefont {Liu}, \citenamefont {Xie},
  \citenamefont {Bi}, \citenamefont {Tan}, \citenamefont {Deng}, \citenamefont
  {Qiu},\ and\ \citenamefont {Peng}}]{liu2020}%
  \BibitemOpen
  \bibfield  {author} {\bibinfo {author} {\bibfnamefont {Z.}~\bibnamefont
  {Liu}}, \bibinfo {author} {\bibfnamefont {K.}~\bibnamefont {Guo}}, \bibinfo
  {author} {\bibfnamefont {G.}~\bibnamefont {Hu}}, \bibinfo {author}
  {\bibfnamefont {Z.}~\bibnamefont {Shi}}, \bibinfo {author} {\bibfnamefont
  {Y.}~\bibnamefont {Li}}, \bibinfo {author} {\bibfnamefont {L.}~\bibnamefont
  {Zhang}}, \bibinfo {author} {\bibfnamefont {H.}~\bibnamefont {Chen}},
  \bibinfo {author} {\bibfnamefont {L.}~\bibnamefont {Zhang}}, \bibinfo
  {author} {\bibfnamefont {P.}~\bibnamefont {Zhou}}, \bibinfo {author}
  {\bibfnamefont {H.}~\bibnamefont {Lu}}, \bibinfo {author} {\bibfnamefont
  {M.-L.}\ \bibnamefont {Lin}}, \bibinfo {author} {\bibfnamefont
  {S.}~\bibnamefont {Liu}}, \bibinfo {author} {\bibfnamefont {Y.}~\bibnamefont
  {Cheng}}, \bibinfo {author} {\bibfnamefont {X.~L.}\ \bibnamefont {Liu}},
  \bibinfo {author} {\bibfnamefont {J.}~\bibnamefont {Xie}}, \bibinfo {author}
  {\bibfnamefont {L.}~\bibnamefont {Bi}}, \bibinfo {author} {\bibfnamefont
  {P.-H.}\ \bibnamefont {Tan}}, \bibinfo {author} {\bibfnamefont
  {L.}~\bibnamefont {Deng}}, \bibinfo {author} {\bibfnamefont {C.-W.}\
  \bibnamefont {Qiu}}, \ and\ \bibinfo {author} {\bibfnamefont
  {B.}~\bibnamefont {Peng}},\ }\href {\doibase 10.1126/sciadv.abc7628}
  {\bibfield  {journal} {\bibinfo  {journal} {Sci. Adv.}\ }\textbf {\bibinfo
  {volume} {6}},\ \bibinfo {pages} {eabc7628} (\bibinfo {year}
  {2020})}\BibitemShut {NoStop}%
\bibitem [{\citenamefont {Shan}\ \emph {et~al.}(2020)\citenamefont {Shan},
  \citenamefont {Bas}, \citenamefont {Lisenkov}, \citenamefont {Matyushov},
  \citenamefont {Sun},\ and\ \citenamefont {Page}}]{shanj2020}%
  \BibitemOpen
  \bibfield  {author} {\bibinfo {author} {\bibfnamefont {P.~J.}\ \bibnamefont
  {Shan}}, \bibinfo {author} {\bibfnamefont {D.~A.}\ \bibnamefont {Bas}},
  \bibinfo {author} {\bibfnamefont {I.}~\bibnamefont {Lisenkov}}, \bibinfo
  {author} {\bibfnamefont {A.}~\bibnamefont {Matyushov}}, \bibinfo {author}
  {\bibfnamefont {N.~X.}\ \bibnamefont {Sun}}, \ and\ \bibinfo {author}
  {\bibfnamefont {M.~R.}\ \bibnamefont {Page}},\ }\href {\doibase
  10.1126/sciadv.abc5648} {\bibfield  {journal} {\bibinfo  {journal} {Sci.
  Adv.}\ }\textbf {\bibinfo {volume} {6}},\ \bibinfo {pages} {eabc5648}
  (\bibinfo {year} {2020})}\BibitemShut {NoStop}%
\bibitem [{tru()}]{truell1969}%
  \BibitemOpen
  \href@noop {} {}\bibinfo {note} {R. Truell, C. Elbaum, and B. B. Chick,
  \emph{Ultrasonic Methods in Solid State Physics} (Academic Press, New York,
  1969).}\BibitemShut {Stop}%
\bibitem [{lut()}]{luthi2004}%
  \BibitemOpen
  \href@noop {} {}\bibinfo {note} {B. L\"uthi, \emph{Physical Acoustics in the
  Solid State} (Springer-Verlag, Berlin, 2004).}\BibitemShut {Stop}%
\end{thebibliography}
\end{document}